\documentclass[a4paper, oneside, twocolumn, notitlepage, 10pt]{extarticle_ecoc}
\usepackage{ecoc}
\usepackage{tw}
\usepackage{caption}
\usepackage{subcaption}
\begin{document}
\selectlanguage{english} 
\title{Experiments on Bipolar Transmission with Direct Detection}%

\author{
    Thomas Wiegart\textsuperscript{(1),$\dagger$},
    Daniel Plabst\textsuperscript{(1)},
    Tobias Prinz\textsuperscript{(1)}, 
    Talha Rahman\textsuperscript{(2)},
    Maximilian Schädler\textsuperscript{(2)},\\
    Neboj\v{s}a Stojanovi\'{c}\textsuperscript{(2)},
    Stefano Calabr\`{o}\textsuperscript{(2)},
    Norbert Hanik\textsuperscript{(1)},
    Gerhard Kramer\textsuperscript{(1)}
}

\maketitle

\begin{strip}
 \begin{author_descr}
    
   \textsuperscript{(1)} Institute for Communications Engineering, Technical University of Munich,  Germany, \textsuperscript{$\dagger$}\textcolor{blue}{\uline{tw@tum.de}}

   \textsuperscript{(2)} Huawei Technologies Duesseldorf GmbH, Munich Research Center, Germany

 \end{author_descr}
\end{strip}

\newcommand{\tw}[1]{\textcolor{red}{#1}}

\setstretch{1.1}

\renewcommand\footnotemark{}
\renewcommand\footnoterule{}

\begin{strip}
  \begin{ecoc_abstract}
    Achievable information rates of bipolar 4- and 8-ary constellations are experimentally compared to those of intensity modulation (IM) when using an oversampled direct detection receiver. The bipolar constellations gain up to $\SI{1.8}{dB}$ over their IM counterparts.
  \end{ecoc_abstract}
\end{strip}

\section{Introduction}
Short-reach fiber-optic communication systems usually use \ac{DD} receivers to reduce hardware cost, system complexity, and energy consumption as compared to coherent receivers~\cite{chagnon_optical_comms_short_reach_2019,zhong_dsp_for_short_reach_2018,qian_imdd_beyond_bw_lim_dcoi_2019}. A \ac{DD} device outputs the intensity of its input signal and thus short-reach systems with \ac{DD} usually apply single polarization \ac{IM}, i.e., information is modulated onto the intensity and the transmit constellation is real and non-negative, e.g., \ac{OOK} or $4$-/$8$-ary \ac{PAM}.

However, the signal phase can be detected with a \ac{DD} receiver by employing oversampling~\cite{mecozzi16}, and when optical noise dominates one loses at most one \ac{bpcu} as compared to coherent detection~\cite{mecozzi18, tasbihi20}. This motivates using bipolar or even complex-valued modulation formats.
Two oversampled receivers for bipolar and complex-valued modulation formats were proposed in~\cite{secondini20,tasbihi21}. They recover the phase from limited \ac{ISI} in the received signal. More sophisticated receivers for bandlimited channels were developed in~\cite{plabst22} and numerical simulations showed significant energy gains of bipolar modulation formats over \ac{IM}.

In this paper, we experimentally demonstrate that bipolar transmission improves IM by using the tools and methods in~\cite{plabst22}. We perform measurements with optical \ac{B2B} transmission and with $\SI{20}{km}$ of \ac{SSMF} in C-band. We observe gains of up to $\SI{1.3}{dB}$ and $\SI{1.8}{dB}$, respectively, for $4$-ary and $8$-ary bipolar constellations as compared to unipolar $4$- and $8$-PAM.
\begin{figure*}[t]
    \centering
    \includegraphics[width=\linewidth]{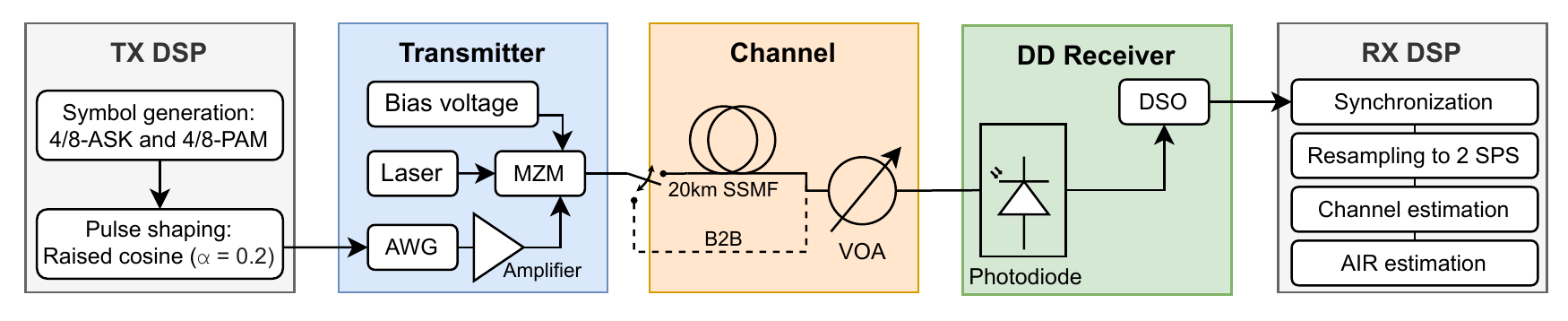}
    \caption{Experimental setup: raised cosine pulses with roll-off $\alpha=0.2$ and $4$- and $8$-ary ASK and PAM constellations.}
    \vspace{0.3cm}
    \label{fig:setup}
\end{figure*}
\tikzset{every picture/.style={line width=0.75pt}}
\begin{figure*}[t]
	\centering
	\begin{subfigure}[b]{0.5\linewidth}
		\includegraphics{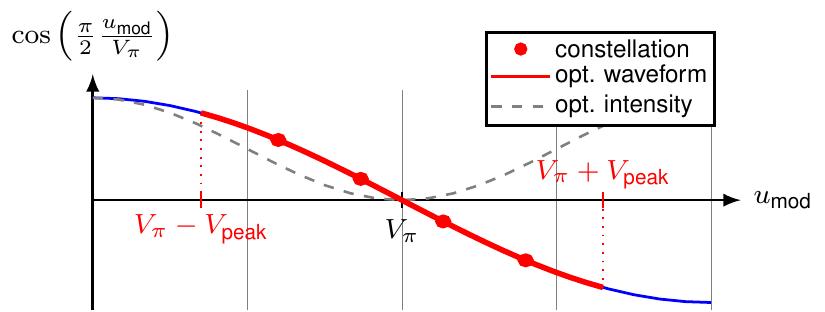}
		\caption{Bipolar $4$-ASK modulation. }
		\label{fig:figa}
	\end{subfigure}\hfill
	\begin{subfigure}[b]{0.5\linewidth}
		\includegraphics{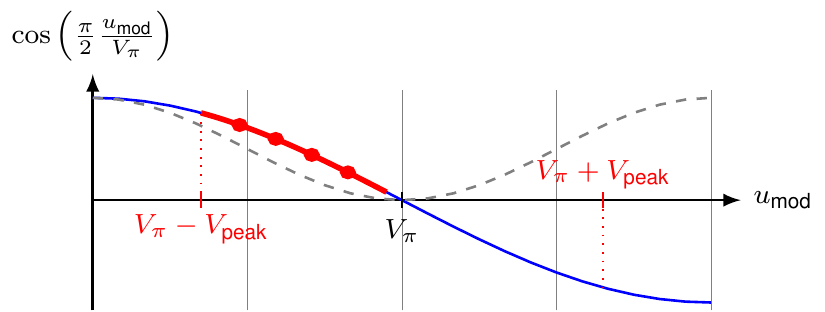}
		\caption{Unipolar PAM modulation with non-negative amplitudes.}
		\label{fig:figb}
	\end{subfigure}
	\vspace{0.2cm}
\caption{Visualization of bipolar and unipolar modulation using a MZM with $4$-ASK and $4$-PAM. Red dots denote constellation points and red lines show all possible values of the waveform. The gray dashed curve depicts the intensity of the modulated signal.}
\end{figure*}
\section{System Model}
We adapt the system model from~\cite{plabst22} and transmit a block of $n$ uniform \ac{i.i.d.} symbols $X_i, \, i \in \{1, \dots, n\}$, from the alphabet $\cX_\text{4-PAM} = \{0, 1, 2,3\}$ for $4$-ary \ac{PAM} or from $\cX_\text{4-ASK} = \{-3, -1, 1, 3 \}$ for $4$-ary \ac{ASK}, and analogously for $8$-PAM and $8$-ASK.
The transmit symbols are pulse shaped, modulated, and transmitted through the channel. A bandlimited \ac{DD} receiver outputs the received signal intensity and the receiver uses two-fold oversampling.

Consider the upsampled transmit sequence $\bm{X'} = [X_1, 0, X_2, 0, \dots, X_n, 0]$ and the channel matrix $\bm{H}$ which has Toeplitz structure and is constructed from a $M$-tap impulse response $\bm{h} = [h_1, \dots, h_M]$. The matrix maps $2n$ inputs to $2n$ outputs, i.e., we discard border samples. The channel matrix includes all filters and linear transmission effects, i.e., pulse shaping, bandwidth limitations, \ac{CD}, and attenuation.

We model the time-discrete oversampled baseband system by:
\vspace{-0.2cm}
\begin{equation} \vspace{-0.2cm}
    \bm{Y} = \left| \bm{H} \bm{X}' + \bm{N}_1  \right|^{\circ 2} + \bm{N}_2 \in \mathbb{R}^{2n}
    \label{eq:model}
\end{equation}
where $|\cdot|^{\circ 2}$ denotes the element-wise modulus squared, and where $\bm{N}_1$ and $\bm{N}_2$ are additive noise before and after the \ac{PD}, respectively.
The entries of $\bm{N}_1$ are modelled as independent circularly-symmetric complex Gaussian (CSCG) and they include noise impairments of the transmitter, e.g., laser phase noise or amplifier noise. The entries of $\bm{N}_2$ are modelled as zero-mean independent real Gaussian and they include noise impairments of the receiver after the \ac{PD} (e.g., thermal noise). Comparing to the model from~\cite{plabst22}, we added the noise $\bm{N}_1$ before the \ac{PD} because we observed significant transmitter noise in our experiments.

\section{Achievable Information Rates}
The \ac{MI} rate $\tfrac1n I(\bm{X}'; \bm{Y})$ is an \ac{AIR} for reliable communication. We follow the algorithm described in~\cite{plabst22} to compute the \ac{MI} for channels with ISI in the oversampled receive sequences. For spectrally-efficient pulses, e.g., raised cosine pulses with small roll-off factors, there is significant ISI.
However, the algorithm in \cite{plabst22} has an exponentially growing complexity in the number of considered ISI taps. We thus compute lower bounds on the MI through an auxiliary channel with $L$ taps where $L$ is smaller than the actual system memory $M$ (see~\cite{arnold06} and Sec.~III in~\cite{plabst22}). The resulting rates can be achieved by mismatched decoding, i.e., the receiver uses the auxiliary model to decode.

\section{Experimental Setup}
The experimental setup is shown in Fig.~\ref{fig:setup}. The transmitter generates symbols at symbol rate $R=\SI{30}{GBd}$ and performs differential precoding as in \cite{plabst22}. A raised cosine pulse with roll-off $\alpha=0.2$ is used and the oversampled signal is loaded into an \ac{AWG} operating at $\SI{120}{GSa/s}$. The \ac{AWG} signal is amplified and fed to a \ac{MZM} that is driven by a C-band laser operating at a wavelength of $\lambda = \SI{1550}{nm}$. The bias point of the \ac{MZM} is tuned depending on the constellation as described in the next section. The optical signal is either transmitted in a \ac{B2B} setup with a \ac{VOA} or over $\SI{20}{km}$ of \ac{SSMF} (attenuation $\SI{0.2}{dB/km}$, group velocity dispersion $D = \SI{17}{ps/(nm \cdot km)}$) followed by a \ac{VOA}. This allows for measurements with and without \ac{CD}.

The receiver has a $\SI{70}{Ghz}$ \ac{PD} followed by a \ac{DSO} operating at $\SI{256}{GSa/s}$. Offline processing is performed for synchronization and resampling to $\SI{2}{SPS}$ (samples per symbol) and channel estimation. Afterwards, the achievable rate is computed using the method from \cite{plabst22} with $10000$ samples per rate point.

\section{Generation of ASK and PAM Constellations}
The experimental comparison of ASK and PAM constellations is performed as follows. For both constellations, the \ac{AWG} is loaded with the same bipolar (mean-free) transmit sequence.

For ASK transmission, the \ac{MZM} is tuned to the null point $V_\pi$ and modulated with a signal with a peak-to-peak voltage span of $2V_\text{peak}$ as depicted in Fig.~\ref{fig:figa}. The output signal thus uses both positive and negative amplitudes.

For PAM transmission, the peak-to-peak output voltage of the \ac{AWG} is halved compared to ASK transmission and the bias point of the \ac{MZM} is adjusted so that the PAM constellation has the same average optical launch power as in the ASK case. For the considered scenario (raised-cosine pulse with $\alpha=0.2$) this also leads to almost the same peak power for both scenarios (ASK had a peak power which is only $\SI{0.05}{dB}$ larger). Thus the two constellations are comparable under average or peak power constraints.
The launch power into the \ac{SSMF} or the \ac{VOA} was $\SI{-3.2}{dBm}$ for $4$-ary constellations and $\SI{-5.0}{dBm}$ for $8$-ary constellations.

\section{Auxiliary Channel Optimization}
To consider additional effects occurring in the experiments, we slightly modify~\eqref{eq:model} to
\begin{equation}
    \bm{\tilde{Y}} = \left| \bm{H} \bm{X}' + \bm{N}_1 + \bm{\mu}_1  \right|^{\circ 2} + \bm{N}_2 + \bm{\mu}_2
    \label{eq:model2}
\end{equation}
to include the MZM bias $\bm{\mu}_1 = (\bm{1}_{n} \otimes [\mu_{11}, \mu_{12}])^\text{T}$, where $\bm{1}_n$ is the length $n$ all-ones vector, $\otimes$ denotes the Kronecker product and $\mu_{11}, \mu_{12}$ are design parameters.  Furthermore, \eqref{eq:model2} includes $\bm{\mu}_2 = (\bm{1}_n \otimes [\mu_{21}, \mu_{22}])^\text{T}$ which models DC-free measurements at the oscilloscope. The model~\eqref{eq:model2} has independent CSCG noise $\bm{N}_1$ and zero-mean real Gaussian noise $\bm{N}_2$ with covariance matrices $\bm{C}_{\bm{N}_1} = \mathrm{diag}(\bm{1}_n \otimes [\sigma^2_{11}, \sigma^2_{12}])$ and $\bm{C}_{\bm{N}_2} = \mathrm{diag}(\bm{1}_n \otimes [\sigma^2_{21}, \sigma^2_{22}])$, respectively. This allows to individually adjust the means $\mu_{ij}$ and the variances $\sigma^2_{ij}$ for samples at and between symbol times.

The parameters $(\mu_{11},\ldots,\mu_{22}, \sigma_{11}^2,\ldots, \sigma^2_{22}, \bm{h})$, including the length $L$ impulse response $\bm{h}$, in the auxiliary model~\eqref{eq:model2} are estimated using the experimental measurements. We remark that these parameters must be carefully chosen to ensure satisfactory performance.
We use the approach described in Sec.~III C of~\cite{plabst22} to optimize the parameters to maximize the \ac{AIR}s. The optimization is carried out numerically using 5000 pilot symbols from the experiments. 
\section{Experimental Results}
\begin{figure}[t!]
    \setlength{\belowcaptionskip}{0.5\baselineskip}
    \begin{subfigure}[c]{\linewidth}
        \includegraphics{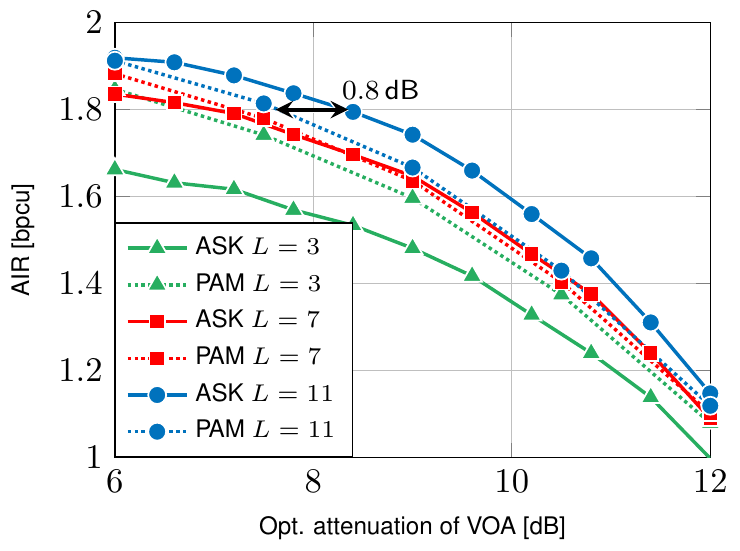}
        \caption{$4$-ASK vs. $4$-PAM in optical B2B setup.}
        \label{fig:4ary_B2B}
    \end{subfigure}
    \begin{subfigure}[t]{\columnwidth}
        \includegraphics{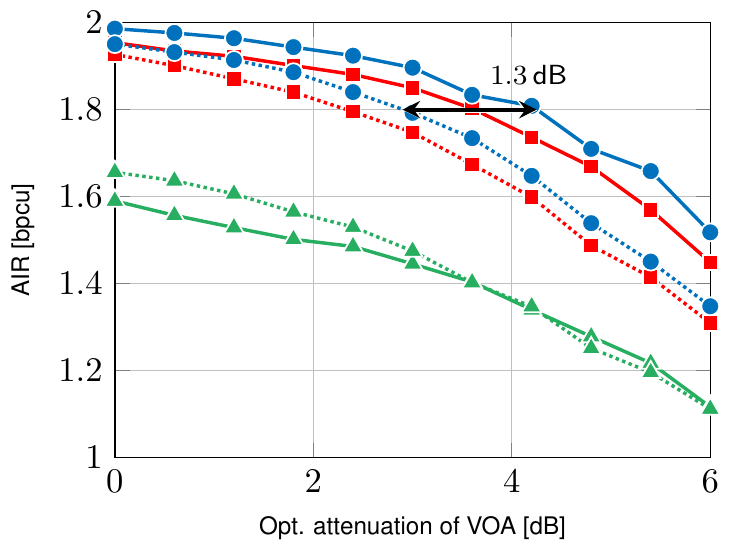}
        \caption{$4$-ASK vs. $4$-PAM with $\SI{20}{km}$ SSMF followed by VOA.}
        \label{fig:4ary_20km}
    \end{subfigure}
    \begin{subfigure}[t]{\columnwidth}
        \includegraphics{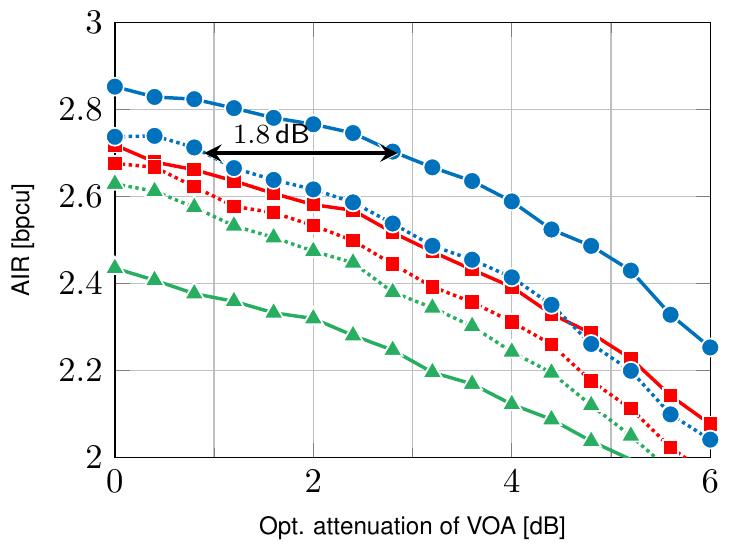}
        \caption{$8$-ASK vs. $8$-PAM in optical B2B setup.}
        \label{fig:8ary_B2B}
    \end{subfigure}
    \caption{AIRs for $Q$-ASK (solid) and $Q$-PAM (dotted) for (a) $Q=4$, optical B2B, (b) $Q=4$ and $\SI{20}{km}$ of SSMF followed by a VOA, and (c) $Q=8$, optical B2B. Colors (marks) correspond to different memory lengths $L$ considered at the receiver, see the legend in subplot (a).}
    \vspace{-0.5cm}
\end{figure}

Fig.~\ref{fig:4ary_B2B} compares AIRs with $4$-ASK and $4$-PAM transmission for different optical attenuation values in a B2B setting. We consider different detector memory values $L$ (note that $L$ taps in the oversampled model correspond to $(L-1)/2$ symbols). PAM significantly outperforms ASK for $L=3$ (1 symbol, green curves). Both schemes perform almost equally well for $L=7$ (3 symbols, red curves).  ASK outperforms PAM by $\SI{0.8}{dB}$ at a rate of $\SI{1.8}{bpcu}$ for $L=11$ (5 symbols, blue curves).
    
The curves in Fig.~\ref{fig:4ary_20km} show results with $\SI{20}{km}$ \ac{SSMF} followed by a \ac{VOA}. As observed in \cite{plabst22}, \ac{CD} may help the receiver to recover phase information and especially ASK benefits from \ac{CD}. In this setting, ASK and PAM perform similarly for $L=3$ and ASK outperforms PAM for larger memory. For $L=11$, ASK gains about $\SI{1.3}{dB}$ over PAM at an AIR of $\SI{1.8}{bpcu}$.
    
Fig.~\ref{fig:8ary_B2B} shows results for $8$-ary transmission in a \ac{B2B} scenario. The results are in line with the $4$-ary ones, but the gains of bipolar transmission increase for higher order constellations. For $L=11$ we see gains of $\SI{1.8}{dB}$ at an \ac{AIR} of $\SI{2.7}{bpcu}$ and an \ac{AIR} gain of $\SI{0.12}{bpcu}$ without attenuation.

For all setups we observe that ASK outperforms PAM if the receiver is capable of handling enough memory. For very low memory, however, PAM might be beneficial.

\section{Conclusions}
We experimentally verified the simulation results of \cite{plabst22}, where bipolar transmission for oversampled \ac{DD} receivers was proposed. We showed gains of up to $\SI{1.8}{dB}$ using bipolar constellations as compared to their unipolar \ac{IM} counterparts.

\section{Acknowledgements}
The authors would like to thank Ullrich Wünsche and Ivan Nicolas Cano Valadez for their help with the lab equipment.

\clearpage
\printbibliography

\vspace{-4mm}

\end{document}